\documentclass[12pt]{article}
\usepackage[colorlinks=true,linkcolor=blue, linktoc=page, urlcolor=blue,citecolor=magenta]{hyperref}
\usepackage{amsmath,amscd}
\usepackage{amsfonts,dsfont}
\usepackage{scalerel}
\usepackage{amssymb}
\usepackage{graphicx,shortvrb}
\usepackage[dvipsnames]{xcolor}
\usepackage{stmaryrd}
\usepackage[utf8]{inputenc}
\usepackage{mdframed}
\usepackage{cite}

\usepackage{wrapfig}

\def\calf{{\cal F}}

\def\calk{{\cal K}}

\def\cale{{\cal E}}

\newcommand{\rp}{\right)}
\newcommand{\lp}{\left(}

\newcommand{\rli}{\right|}

\newcommand{\pr}{\partial}

\DeclareMathOperator{\sgn}{sgn}

\DeclareMathOperator{\R}{\mathbb{R}}

\newcommand{\sgd}{\, .}

\newcommand{\gag}{\quad \mbox{and} \quad}
\newcommand{\gwg}{\quad \mbox{where} \quad}

\newcommand{\emb}{\mathrm{j}}
\newcommand{\U}{\Upsilon}

\usepackage[normalem]{ulem}

\begin{document}

\begin{titlepage}
 \vskip 1.8 cm

\begin{center}{\huge \bf Geodesic motion on the group of boundary diffeomorphisms from Einstein's equations}\\

\end{center}
\vskip .3cm
\vskip 1.5cm

\centerline{\large {{\bf Emine Şeyma Kutluk$^{1}$ and  Dieter Van den Bleeken$^{2,3}$}}}

\vskip 1.0cm

\begin{center} 

    1) Middle East Technical University\\
    06800 Çankaya / Ankara, TURKEY
	
	\vskip 1cm

	2) Primary address:\\
	Physics Department, Boğaziçi University\\
	34342 Bebek / Istanbul, TURKEY
	
	\vskip 1cm
	
	3) Secondary address:\\
	Institute for Theoretical Physics, KU Leuven\\
	3001 Leuven, Belgium
	
	\vskip 1cm
	
\texttt{ekutluk@metu.edu.tr, dieter.van@boun.edu.tr}
\end{center}
\vskip 1.3cm \centerline{\bf Abstract} \vskip 0.2cm \noindent In \cite{Kutluk:2019ghr} it was shown how in an adiabatic limit the vacuum Einstein equations on a compact spatial region can be re-expressed as geodesic equations on the group of diffeomorphisms of the boundary. This is reminiscent of the program initiated by V. Arnold to reformulate models of continuum mechanics in terms of geodesic motion on diffeomorphism groups. We revisit some of the results of \cite{Kutluk:2019ghr} in this light, pointing out parallels and differences with the typical examples in geometric continuum mechanics. We work out the case of 2 spatial dimensions in some detail.

\end{titlepage}
\tableofcontents

\section{Introduction and overview}
One of the most beautiful and elegant observations of modern mathematical physics is that various differential equations of physical interest are equivalent to geodesic equations on Lie groups equipped with a left or right invariant metric, see e.g. \cite{holm2009geometric} for an introduction. The simplest and most widely known result is that the free motion of a rigid body can be interpreted as motion along a geodesic on SO(3), equipped with a metric determined by the moment of inertia of the body. In \cite{Arnold66} V. Arnold generalized this idea, by reformulating the dynamics of an ideal fluid as geodesic motion on the group of volume preserving diffeomorphisms. Since then a variety of similar equivalences have been discovered, for example for Burger's equation, the Camassa-Holm equation and others, see e.g. \cite{arnold1998topological}. By today this has developed into an interesting subfield at the intersection of geometric (continuum) mechanics and the theory of (infinite dimensional) Lie groups.

Independent of the aforementioned work, in \cite{Kutluk:2019ghr} we observed -- together with A. Seraj --  that in an adiabatic\footnote{This slow motion approximation is alternatively referred to as the Manton or moduli space approximation. It is best known through its application in the physics of solitons \cite{Manton:1981mp}. We use the nomenclature of Stuart \cite{Stuart:2007zz}.} approximation the vacuum Einstein equations on a spacetime manifold of the form $\mathbb{R}\times \bar M$ reduce to geodesic motion on Diff$(\partial \bar M)$, the group of boundary diffeomorpisms\footnote{More precisely the motion takes place on the coset Diff$(\partial \bar M)/$ISO$(\partial \bar M)$, but for a generic boundary ISO$(\partial \bar M)=\{\mathrm{id}\}$.}.

It is the aim of this paper to put the results of \cite{Kutluk:2019ghr} into the context of Arnold's program in geometric continuum mechanics. We will use some of the results and language of this field to clarify certain points of \cite{Kutluk:2019ghr}, while also highlighting some differences with the class of examples considered so far in geometric mechanics, such as the pseudo-Riemannian nature of the metric on the diffeomorphism group in our setup. Furthermore, as we will see, the right invariant geometry induced on the diffeomorphism group by the dynamics of general relativity is much more intricate than those in examples so far considered in geometric continuum mechanics.

We begin our paper in Section \ref{sec:diffg}, by reviewing some key mathematical concepts and results concerning geodesic motion on diffeomorphism groups and the corresponding physical dynamics. We present the geodesic equation in a fully covariant form -- a first in the literature as far as we are aware -- and shortly present the classic examples.

In Section \ref{sec:adGR} we review the key results of \cite{Kutluk:2019ghr} expressing them in a mathematically more elegant form and make the connection with the formalism introduced in Section \ref{sec:diffg}. The general procedure is then illustrated in the case of two spatial dimensions, whose detailed treatment is new. This simplest case already shows the intricate nature of the construction and allows us to point out some remaining technical challenges. We end the paper with a short discussion in Section \ref{sec:discuss}.\\

{\bf Note: }Our level of rigor when dealing with infinite dimensional analysis will be that of the typical physics texts in this field, e.g. \cite{holm2009geometric}, trusting on the expert mathematical literare, e.g. \cite{michor1980manifolds, marsden1974applications}, to fill some gaps. We are also somewhat sloppy in our nomenclature. For example we will assume all groups to be connected to the identity, so that what we call the `diffeomorphism group' is in more detailed and standard nomenclature the `identity component of the diffeomorphism group', we'll refer to certain anti-isomorphisms simply as isomorphisms etc. We hope this will ease reading and it will allow us to focus on the main points.

\section{Diffeomorphism groups: right invariant metrics and their geodesics}\label{sec:diffg}
Although our main interest will lie with Diff($M$), the group of diffeomorphisms of a manifold $M$ to itself, some parts of the construction are most clearly exhibited in the slightly larger context of maps between two manifolds $M$ and $N$. With this motivation we introduce the right invariant metric on the manifold of embeddings of $M$ into $N$, before specializing to the case $M=N$, where these embeddings become the diffeomorphisms of $M$. After we have reviewed this standard construction of right invariant metrics on the group of diffeomorphisms we recall the derivation of the corresponding geodesic equation. We end the section by illustrating how for certain choices of diffeomorphism groups and metrics this equation takes the form of some well known equations in continuum mechanics.

Our treatment draws from \cite{michor2006some, holm2009geometric, michor2020manifolds}.

\subsection{Metrics on infinite dimensional manifolds of maps}
\subsubsection{The larger perspective: the manifold of embeddings}
The set $\mathrm{Emb}(M,N)$ of embeddings of $M$ into $N$ can naturally be given the structure of an infinite dimensional manifold. In addition there is a canonical right action of Diff$(M)$ on this manifold, defined via the composition of maps. Indeed, let $f: M\rightarrow N$ be such an embedding and $\phi: M\rightarrow M$ a diffeomorphism, then
\begin{equation}
R_\phi f=f\circ \phi \ .\label{rightaction}
\end{equation}
Although this will play no role in our discussion, it is interesting to point out that this right action makes  $\mathrm{Emb}(M,N)$ a principal fibre bundle with structure group Diff$(M)$. 

A vector tangent to $\mathrm{Emb}(M,N)$  can be obtained by considering a curve in $\mathrm{Emb}(M,N)$, i.e. a one-parameter family of embeddings $f_s$. The vector is then identified with the derivative $\dot f=\frac{df_s}{ds}$. To understand this object better, note that given $p\in M$ then $f_s(p)$ defines a curve in $N$ and so $\dot f(p)=\frac{d}{ds}f_s(p)$ defines a tangent vector in $T_{f(p)}N$. So we see that the tangent vector $\dot f$ is a map from $M$ into the tangent bundle of $N$. In more mathematical term; a tangent vector to $\mathrm{Emb}(M,N)$ at $f$ is a section of the pull-back bundle\footnote{Note that we can formally identify $M$ with its image $f(M)$ which is a sub-manifold of $N$, we can then think of $f^*TN$ as the restriction of $TN$ to this sub-manifold.} $f^* TN$: 
\begin{equation}
T_f\mathrm{Emb}(M,N)=\Gamma_c(f^*TN) \ .\label{tangemb}
\end{equation}

Given $g$, a Riemannian metric on $N$, there exists an associated metric $G$ on $\mathrm{Emb}(M,N)$. First one observes that via the discussion above $g_p(\dot f_1(p),\dot f_2(p))$ is well defined for all $p\in M$ and $\dot f_1,\dot f_2\in T_f\mathrm{Emb}(M,N)$. One then obtains a metric $G$ on $\mathrm{Emb}(M,N)$ by integrating over $M$:
\begin{equation}
G_f(\dot f_1,\dot f_2)=\int_M g(\dot f_1,\dot f_2)\,\mathrm{vol} f^*g \ .\label{infmetric}
\end{equation}
Here $\mathrm{vol} f^*g$ is the volume element on $M$ determined by the induced metric $f^*g$ on $M$, the pull-back by $f$ of the metric $g$ on $N$.

Since integration over $M$ is diffeomorphism invariant, one finds that the metric \eqref{infmetric} is invariant under the right-action \eqref{rightaction}:
\begin{equation}
G_{R_\phi f}(dR_\phi(\dot f_1),dR_\phi(\dot f_2))=G_f(\dot f_1,\dot f_2) \ ,\label{rightinv}
\end{equation}
where one uses that $(dR_\phi \dot f)(p)=\dot f (\phi(p))$ and $\mathrm{vol}(R_\phi f)^*g=\phi^*\mathrm{vol} f^*g$.

\subsubsection{The case of interest: the group of diffeomorphisms}
If we specialize now to the case $N=M$ one obtains $\mathrm{Emb}(M,M)=\mathrm{Diff}(M)$, which naturally is an infinite dimensional Lie group. Letting $\mathrm{id}$ denote the identity map we observe that \eqref{tangemb} becomes
\begin{equation}
T_\mathrm{id} \mathrm{Diff}(M)=\Gamma_c(TM)=\mathfrak{X}(M) \ .
\end{equation}
That is, we recover the well-known fact that the Lie-algebra associated to diffeomorphisms of a manifold $M$ is naturally isomorphic to the algebra of vectorfields on $M$. Let us stress that generic tangent vectors $\dot \phi\in T_\phi \mathrm{Diff}(M)$ are {\it not} in general vector fields on $M$, rather they are sections of the pullback bundle $\phi^* TM$, per \eqref{tangemb}.

Using the right invariance \eqref{rightinv} one can relate the metric \eqref{infmetric} on $\mathrm{Diff}(M)$ to an inner product on $\mathfrak{X}(M)$:
\begin{equation}
G_\phi (\dot \phi_1,\dot \phi_2)=( dR_{\phi^{-1}}\dot\phi_1,dR_{\phi^{-1}}\dot\phi_2)\label{candifmet}
\end{equation}
where
\begin{equation}
( \chi_1,\chi_2)=\int_M g(\chi_1,\chi_2)\,\mathrm{vol}\,  g\quad\mbox{with}\quad \chi_1,\chi_2\in \mathfrak{X}(M) \ .
\end{equation}
This reveals the option to define a more general class of inner products, and also metrics, as follows. First one defines
\begin{equation}
\langle \chi_1,\chi_2\rangle=(\chi_1,\mathbb{D}\chi_2)=\int_M g(\chi_1,\mathbb{D}\chi_2)\,\mathrm{vol}\,  g \ .\label{inprod}
\end{equation}
This is a proper inner product when $\mathbb{D}$ is a positive definite, self-adjoint operator on the Hilbert space $\mathfrak{X}(M), (\cdot,\cdot)$. Replacing the canonical inner product $(\cdot,\cdot)$ with the modified one  in \eqref{candifmet}, one obtains a new metric on $\mathrm{Diff}(M)$:
\begin{equation}
G_\phi^\mathbb{D}(\dot \phi_1,\dot \phi_2)=\langle dR_{\phi^{-1}}\dot\phi_1,dR_{\phi^{-1}}\dot\phi_2 \rangle \ .\label{Dmet}
\end{equation}
Note that also this new `deformed' metric $G^\mathbb{D}$ is right invariant by construction.

\subsection{The geodesic equation on the group of diffeomorphisms}

\subsubsection{Derivation of the geodesic equation}
Given the construction of the right invariant metrics \eqref{Dmet} on the group of geodesics of a manifold $M$ it is natural to study the geodesics associated to such a metric. We'll take a somewhat pedestrian approach inspired by mechanics, and will define the geodesics associated to a metric $G^\mathbb{D}$ to be those curves $\phi(t)$ that extremize the action
\begin{equation}
S=\int \frac{1}{2}G^\mathbb{D}_{\phi(t)}(\dot \phi (t),\dot \phi(t))dt \ ; \label{action1}
\end{equation}
where $\dot \phi(t)=\frac{d}{dt}\phi(t)$, and, as in the previous subsections, $\phi(t)$ is a diffeomorphism from $M$ to itself for all $t$.

If we choose coordinates $x^i$ on $M$ then we can think of the curve $\phi(t)$ as a one-parameter family of maps $\phi^i(x;t)$. Defining
\begin{equation}
\chi^i(x;t)=(dR_{\phi^{-1}}\dot{\phi})^i(x;t)=\partial_t\phi^i(\phi^{-1}(x;t);t)\label{chidef}
\end{equation}
one sees, via \eqref{Dmet}, that \eqref{action1} becomes
\begin{equation}
S=\int \frac{1}{2}\langle \chi(t),\chi(t)\rangle dt \ . \label{action2}
\end{equation}

In physical terms one can think of $\phi^i(x;t)$ as defining a fluid flow, where $\phi^i(x,t)$ is the position at time $t$ of a particle that at time $t=0$ was at $x^i$ (assuming the convention $\phi^i(x;0)=x^i$). The Lagrangian velocity, i.e. the velocity at time $t$ of the particle which was at $x^i$ at time $t=0$, of the flow is then $\partial_t \phi^i(x;t)$. In this picture $\chi^i(x;t)$ has the interpretation of the Eulerian velocity of the flow, i.e. the velocity at time $t$ of the particle which is at $x^i$ at that time  \cite{holm2009geometric}.

In mathematical terms $\partial_t\phi^i(x;t) \in T_\phi \mathrm{Diff}(M)$, and it gets mapped into $\chi^i(x;t)\in T_{\mathrm{id}} \mathrm{Diff}(M)$ by the right action $dR_{\phi^{-1}}$. This is a coordinate representation of the concepts we introduced in the previous subsections. As we stressed there, $\partial_t\phi^i(x;t)$ is {\it not} a vectorfield on $M$, while $\chi^i(x;t)$ is. Equivalently $\chi$ is an element of the Lie algebra $\mathfrak{diff}(M)$ (while $\partial_t\phi$ is not).

The procedure of rewriting of \eqref{action1} in terms of \eqref{action2} together with the rigorous derivation of the corresponding Euler-Lagrange equations, for an arbitrary Lie group, is known as {\it Euler-Poincar\'e} reduction, see e.g. \cite{holm2009geometric}. Applied to this setting, the starting point is the observation that the form \eqref{action2} of the action implies that a  generic variation $\delta \phi^i$ will only affect the action through $\delta \chi^i$. A short computation reveals the relation between the two variations:
\begin{eqnarray}
\delta\chi^i&=&\partial_t \delta\phi^i(\phi^{-1})+(\partial_j\partial_t\phi^i)(\phi^{-1})\,\delta (\phi^{-1})^j\label{var1}\\
&=&\partial_t\delta \Phi^i+L_\chi\delta\Phi^i\ .\label{var}
\end{eqnarray}
In the last expression we used the notation
\begin{equation}
\delta\Phi^i(x;t)=\delta\phi^i(\phi^{-1}(x;t);t) \ ,
\end{equation}
as well as the Lie derivative
\begin{equation}
L_\chi \delta \Phi^i=\chi^j\partial_j\delta\Phi^i-\delta\Phi^j\partial_j\chi^i \ .
\end{equation}
To obtain \eqref{var} from \eqref{var1} one can use the relations
\begin{eqnarray}
(\partial_j\partial_t\phi^i)(\phi^{-1}(x;t);t)&=&-\partial_k\chi^i(x;t)\,\partial_j\phi^k(\phi^{-1}(x;t);t) \ ,\nonumber\\
\delta(\phi^{-1})^i(x;t)&=&-\partial_j(\phi^{-1})^i(x;t)\,\delta \Phi^i(x;t)\ , \nonumber\\
\partial_t(\phi^{-1})^i(x;t)&=&-\partial_j(\phi^{-1})^i(x;t)\,\chi^i(x;t) \ ,\\
\delta^i_j\nonumber&=&\partial_k\phi^i(\phi^{-1}(x;t);t)\partial_j(\phi^{-1})^k(x;t) \ ,\\
\partial_t\delta\Phi^i(x;t)&=&(\partial_t\delta\phi^i)(\phi^{-1}(x;t);t)+\partial_t(\phi^{-1})^j(x;t)\partial_j\delta\Phi^i(x;t)\nonumber \ .
\end{eqnarray}  
The variation of the action then becomes
\begin{eqnarray}
\delta S&=&\int \langle\chi,\partial_t\delta\Phi^i+L_\chi\delta\Phi^i\rangle dt\\
&=&\int \langle-\partial_t\chi+L^\dagger_\chi \chi,\delta\Phi^i\rangle dt \ .
\end{eqnarray}
One of the key results of the theory of Euler-Poincar\'e reduction, which we will use without further argument, is that a generic variation $\delta \Phi^i$, can be obtained from some variation $\delta \phi^i$. Since the inner product is non-degenerate and $\delta \Phi^i$ can be a generic vector one finds that a stationary action implies
\begin{equation}
\partial_t\chi-L_\chi^\dagger \chi=0 \ .
\end{equation}
This is then the geodesic equation on $\mathrm{Diff}(M)$. Note that this is a first order equation, since it is an equation for the velocity $\chi(t)$, to find the actual corresponding curve of diffeomorphisms $\phi(t)$ one has to additionally integrate \eqref{chidef}. 

A short computation, using the definitions of the inner product, reveals that
\begin{equation}
L_{\chi_1}^\dagger\chi_2=-\mathbb{D}^{-1}\mathbb{L}_{\chi_1}\mathbb{D}\chi_2
\end{equation}
where
\begin{equation}
(\mathbb{L}_{\chi_1}\chi_2)^i=L_{\chi_1}\chi_2^i+(g^{ik}L_{\chi_1}g_{kj}+\delta^i_j\,g^{-1}L_{\chi_1}g)\chi_2^j \ .
\end{equation}
This implies we can rewrite the geodesic equation equivalently as
\begin{equation}
\partial_t\mathbb{D}\chi+\mathbb{L}_\chi \mathbb{D}\chi=0 \ .
\end{equation}
Writing this out explicitly one gets
\begin{equation}
(\partial_t+\chi^j\nabla_j)\mathbb{D}\chi^i+(\nabla^{i}\chi_j+\delta^i_j\nabla_k\chi^k)\mathbb{D}\chi^j=0 \ .\label{geoeq_expl}
\end{equation}
This equation provides a covariant version of the geodesic equation as derived in e.g. \cite{holm2009geometric}.

\subsubsection{Examples appearing in continuum mechanics}\label{sec:ex}
There are a number of physical continuum systems whose mechanical dynamics is described by a geodesic equation on a diffeomorphism group, i.e. a special case of \eqref{geoeq_expl}. In this subsection we shortly mention some of the most well known. In the next section we explain how the Einstein equations in an adiabatic limit can be added to this list.

\begin{itemize}
	\item {\bf Volume preserving diffeomorphisms and Euler's equation}\\
If we restrict to the subgroup of volume preserving diffeomorphisms, such that $L_\chi g=\nabla_i\chi^i=0$, and restrict to the canonical case $\mathbb{D}=\mathds{1}$, then \eqref{geoeq_expl} reduces to
\begin{equation}
\partial_t\chi^i+\chi^j\nabla_j\chi^i+\chi^j\nabla^i\chi_j=0 \ .
\end{equation}
This equation is equivalent to Euler's equation for the (free) motion of an incompressible fluid if we identify $\chi^i$ with the (Eulerian) velocity of the fluid. Introducing the material derivative $D_t=\partial_t+\chi^j\nabla_j$ and the kinetic energy of the fluid $ T=\frac{1}{2}g_{ij}\chi^i\chi^j$ the equation takes the perhaps more familiar form
\begin{equation}
D_t\chi^i=-\nabla^i T \ .
\end{equation}

\item{\bf 1d diffeomorphisms and Burger's equation}\\
If we take $M$ to be one-dimensional and consider the canonical choice $\mathbb{D}=\mathds{1}$, then the geodesic equation \eqref{geoeq_expl} takes the form
\begin{equation}
\partial_t \chi^x+3\chi^x\nabla_x\chi^x=0 \label{burgeq}
\end{equation}
which is the (inviscid) Burger equation\footnote{Writing $u=\chi^x$ and assuming the coordinate $x$ to be Cartesian, so that $\nabla_x=\partial_x$, reproduces the standard form $\partial_t u+3 u \partial_xu=0$.}.

\item{\bf 1d diffeomorphisms and the Camassa-Holm equation}\\
When one considers one-dimensional diffeomorphisms, but now with a metric based on the operator\footnote{Note that when $\mathbb{D}=\mathds{1}-\nabla_x^2$ then the inner product \eqref{inprod} on $\mathfrak{X}(M)$ reproduces the $H^1$ Sobolev norm:
	\begin{equation*}
	\langle \chi,\chi\rangle=\int_M \sqrt{g}(\chi^2+(\nabla\chi)^2)dx=\|\chi\|^2_{H^1} \ .
	\end{equation*}} $\mathbb{D}=\mathds{1}-\nabla_x^2$, the geodesic equation \eqref{geoeq_expl} becomes
\begin{equation}
\partial_t \chi^x-\partial_t\nabla_x^2\chi^x+3\chi^x\nabla_x\chi^x-2\nabla_x\chi^x \nabla_x^2\chi^x-\chi^x\nabla_x^3\chi^x=0 \ .
\end{equation}
This equation is the (dispersionless) Camassa-Holm equation\footnote{To get a more familiar form one can again write $u$ for the vector component $\chi^x$, and assume $x$ to be Cartesian so that $\nabla_x=\partial_x$.} describing shallow water waves, see e.g. \cite{holm2009geometric}.

\end{itemize}

\section{From General Relativity to geodesics on the group of boundary diffeomorphisms}\label{sec:adGR}
In this section we review some of the key results of \cite{Kutluk:2019ghr}. To be compatible with the previous section we change some of the notation and choice of symbols that were used in \cite{Kutluk:2019ghr}, see Appendix \ref{notapp} for a dictionary. 

First we introduce the operator $\mathbb{D}$ that via (\ref{inprod}, \ref{Dmet}) defines the right invariant metric on the diffeomorphism group. Only afterwards, in Section \ref{subsec:sup}, will we explain how this particular metric/operator follows from the vacuum Einstein equations. As an illustration,  in the last subsection we then explicitly compute $\mathbb{D}$ in the simple case of $\dim M=1$.

With respect to \cite{Kutluk:2019ghr} the elegant form \eqref{Ddef} of $\mathbb{D}$ is new, as well as the results in Section \ref{2dsec}.

\subsection{An extrinsic metric on the group of boundary diffeomorphisms}
Let $\bar M$ be a compact subset of $\mathbb{R}^{n+1}$, for simplicity assumed to be homeomorphic to the closed $(n+1)$-ball. We'll denote it's boundary as $M=\partial \bar M$ and write $\emb: M \rightarrow \bar M$ for the related embedding. The flat Euclidean metric $\bar g_\mathrm{o}$ on $\bar M$ naturally induces a metric $g=\emb^*\bar g_\mathrm{o}$ on $M$. Abstractly $\bar M$ is a manifold with boundary which allows us to borrow some machinery from the general theory of manifolds with boundary, in particular related to boundary value problems for differential forms, following\footnote{See also Appendix C in \cite{Kutluk:2019ghr} for a summary of the parts most relevant to our discussion.} \cite{Schwarz:1995}. We expect much of the discussion below to generalize to generic Riemannian manifolds with boundary of general topology. 

Our aim is to introduce an operator $\mathbb{D}$ on the Hilbert space $ \mathfrak{X}(M), (\cdot,\cdot) $, which in turn -- by the discussion of Section \ref{sec:diffg} -- defines a right invariant metric of the type (\ref{inprod}, \ref{Dmet}) on $\mathrm{Diff}(M)$. What is crucial is that in the construction of $\mathbb{D}$ we will make use of the `bulk' space $\bar{M}$ of which $M$ is the boundary, i.e. $\mathbb{D}$ is {\it extrinsic} to $M$, rather than intrinsic.

We will now specify the operator $\mathbb{D}$ relevant to us. We start by considering the map
\begin{equation}
\bar\theta : \Lambda^1(M)\rightarrow \Lambda^2(\bar M) : \omega \mapsto \bar{\theta}_\omega \ ,\label{thetamap}
\end{equation}
defined such that
\begin{equation}
\delta \bar\theta_\omega=d\bar\theta_\omega=0\,,\qquad \emb^*\bar\theta_\omega=d\omega\,.\label{thetaeqs}
\end{equation}
Note that $\bar{\theta}_\omega$ is a {\it harmonic field}, i.e. a form that is both closed and co-closed -- which on a manifold with boundary is a stronger condition than just being a harmonic form. Given the boundary 1-form $\omega$, the bulk two form $\bar\theta_\omega$, defined through the boundary value problem above, is guaranteed to exist and be unique\footnote{Uniqueness is guaranteed since, due its trivial topology, there are no non-zero Dirichlet fields -- harmonic fields whose pullback on the boundary vanishes -- on $\bar M$, except when $\dim \bar M=2$. The case $\dim \bar M=2$ needs to be treated separately and one needs to change the boundary condition on $\bar{\theta}_\omega$ -- since $d\omega=0$ trivially on the one dimensional boundary. In Section \ref{2dsec} we will discuss the case $\dim \bar M=2$ in detail, see \eqref{2dbndc} for the modified boundary value problem defining $\bar{\theta}_\omega$ in that case.}, see e.g. \cite{Schwarz:1995} Thm 3.2.5. 

Using the linear map \eqref{thetamap} we then define
\begin{equation}
\mathbb{D}\,:\, \mathfrak{X}(M)\rightarrow\mathfrak{X}(M)\,:\, \chi \mapsto (\emb^* (\imath_n \bar{\theta}_{\chi^\flat})-2\imath_\chi K)^\sharp\  .\label{Ddef}
\end{equation}
Here $\imath_n$ is the contraction with $n$, a vectorfield whose restriction to $M$ is the unit outward normal vector field to $M$. We have used the musical isomorphism associated to the metric $g$; i.e. $\flat\,:\, TM\rightarrow T^*M\, :\, \chi \mapsto \chi^\flat=g(\cdot,\chi)$ with inverse $\sharp :\, T^*M\rightarrow TM\, : \omega \mapsto \omega^\sharp$. These are mathematical representations of what most physicists respectively refer to as lowering and raising indices. 

In the second term we have introduced the extrinsic curvature $K=\frac{1}{2}\emb^*(L_n\bar g_\mathrm{o})$ of the boundary as a hypersurface inside $(\bar M,\bar g_{\mathrm{o}}).$ Note that $K$ is naturally a symmetric bilinear form on $TM$ and so $\imath_\chi K$ defines a one-form on $M$ for every vectorfield $\chi$ on $M$.

One can show, see the next subsection for an argument, that $\mathbb{D}$ is symmetric with respect to the inner product
\begin{equation}
(\chi_1,\chi_2)=\int_M g(\chi_1,\chi_2)\,\mathrm{vol}\,g \ .
\end{equation}
The associated right invariant metric on $\mathrm{Diff}(M)$ is then defined as in \eqref{Dmet}:
\begin{equation}
G_\phi^\mathbb{D}(\dot\phi_1,\dot \phi_2)=(\chi_1,\mathbb{D}\chi_2)\label{GRmetric}
\end{equation}
where
\begin{equation}
\qquad \chi=dR_{\phi^{-1}}\dot \phi\,,\mbox{ i.e. } \chi^i(x;t)=\partial_t\phi^i(\phi^{-1}(x,t);t) \ .
\end{equation}
As we will also discuss further in the next subsection, the operator $\mathbb{D}$ defined in \eqref{Ddef} is however not positive, so that $G$ is actually a {\it pseudo-Riemannian} metric on $\mathrm{Diff}(M)$.

\subsection{From superspace to boundary diffeomorphisms}\label{subsec:sup}
We now review how the geodesic motion on the group of boundary diffeomorphisms -- equipped with the right invariant metric \eqref{GRmetric} which itself is determined by the operator \eqref{Ddef} -- is equivalent to the Einstein equations in an adiabatic limit. We only summarize the main points and refer to \cite{Kutluk:2019ghr} for a detailed argumentation. In this subsection we switch to the index/component notation more often used in physics and general relativity (GR). 

Before providing some concrete technical details in Section \ref{WdWsection}, let us summarize the main idea in words.

In the geometrodynamic picture, see e.g. \cite{Giulini:1993ct} for a short review, the vacuum Einstein equations can be interpreted as the motion of a particle on superspace -- the space of spatial Riemannian metrics -- equipped with the Wheeler-deWitt metric. The Euler-Lagrange equations of such a particle lead to a geodesic equation on superspace, but one that is sourced by forces arising from a non-trivial potential on superspace -- set by the curvature of the spatial metric. In a slow velocity approximation, introduced by Manton \cite{Manton:1981mp}, one can restrict motion to the subspace of zero potential -- referred to as {\it the space of vacua} in \cite{Kutluk:2019ghr} -- where the motion becomes free, i.e. purely geodesic with respect to the induced metric. This very general argument thus suggests that, just like any system described by a natural Lagrangian \cite{arnol2013mathematical}, the Einstein equations can, in an adiabatic limit, be reduced to purely geodesic motion. At first sight one could be tempted to think the setup is trivial, since vanishing potential implies vanishing Ricci curvature of the 3d spatial metric, and there is, up to diffeomorphism, only one such metric, the flat one. The subtlety is however that in the presence of a boundary\footnote{In this note and in \cite{Kutluk:2019ghr} we focus on the case where space is a compact region with boundary. On a non-compact space the boundary diffeomorphisms should be replaced by the so called asymptotic symmetries -- see e.g. \cite{Barnich:2001jy} or \cite{strominger2017lectures} more recently -- , but it remains an open problem to understand if some of the methods of \cite{Kutluk:2019ghr} can be applied to that setting as well.} not all diffeomorphisms should be thought of as redundancies describing the metrics that are physically the same. Those diffeomorphisms that do not vanish on the boundary will generate physically inequivalent metrics. From this observation it follows, see \cite{Kutluk:2019ghr} for a few extra details, that the zero-potential surface in the superspace of a manifold with boundary can be identified with the group of boundary diffeomorphisms. The metric describing the geodesic motion of interest is then nothing but the pull-back of the Wheeler-deWit metric to this surface.

\subsubsection{Pulling back the Wheeler-deWitt metric}\label{WdWsection}
Let us now make this concrete. Consider the following\footnote{Locally any Lorentzian metric can be brought in this form, these are so called Gaussian normal coordinates.} Lorentzian metric on $\mathbb{R}\times \bar M$
\begin{equation}
ds^2=-dt^2+\bar g_{ij}(x;t)dx^idx^j \ .\label{lormet}
\end{equation}
We will furthermore assume\footnote{This is the restriction to the zero-potential surface in superspace.} the spatial metric $\bar g_{ij}$ to be flat at all times $t$ -- note that this does {\it not} imply that the $n+2$ dimensional metric \eqref{lormet} is flat.
The vacuum Einstein equations for the Lorentzian metric \eqref{lormet} then take the form
\begin{eqnarray}
\bar\nabla^i \left( \dot {\bar g}_{ij}- \bar g^{kl}\dot{\bar g}_{kl} \bar g_{ij} \right) &=&0 \, , \label{const1} \\
\frac{1}{2}\bar g^{ij}\bar g^{kl}\dot {\bar g}_{i[j}\dot {\bar g}_{k]l}&=&0\label{const2} \, ,\\
\ddot{\bar g}_{ij} + \bar g^{kl}(\dot{\bar g}_{ik} \dot{\bar g}_{jl}-\dfrac{1}{2}\dot{\bar g}_{ij} \dot{\bar g}_{kl})&=&0 \, . \label{dynamicaleq}
\end{eqnarray}
The first two equations are constraints while the last equation is a dynamical equation that can be obtained from extremizing the Lagrangian
\begin{eqnarray}
L=\int \frac{1}{2}\bar{\mathrm{G}}_{\bar g}(\dot{\bar g},\dot{\bar g})dt \ .
\end{eqnarray}
This is a purely kinetic term based on the Wheeler-de Witt metric
\begin{equation}
\bar{\mathrm{G}}_{\bar g}(\dot{\bar g}_1,\dot{\bar g}_2)=\frac{1}{2}\int_{\bar M}\bar g^{i[k}g^{j]l}\dot{\bar g}_{1\,{ij}}\dot{\bar g}_{2\,{kl}}\,\sqrt{\mathrm{det}\bar g}\, d^{n+1}x \ .\label{wdw}
\end{equation}
The key point is now that since $\bar g_{ij}$ is flat for all $t$ we can write
\begin{equation}
\bar g(t)=\bar\phi^*(t)\bar g_{\mathrm{o}}\,,
\end{equation}
i.e. the time dependent metric is simply the pull-back by a time dependent diffeomorphism of a fixed, i.e. time-independent, reference flat metric $\bar g_\mathrm{o}$. One can then compute\footnote{See e.g. Appendix A in \cite{Kutluk:2019ghr}.} that
\begin{equation}
\dot{\bar g}=\bar\phi^*(L_{\bar\chi} \bar g_\mathrm{o})\,,\quad\mbox{where}\quad \bar\chi=\dot{\bar{\phi}}\circ\bar \phi^{-1}=dR_{\bar\phi^{-1}}\dot{\bar{\phi}} \ .\label{veldef}
\end{equation}
Here the diffeomorphism $\bar{\phi}$ of $\bar M$ is assumed to preserve the boundary $M=\partial \bar M$, i.e. $\bar{\phi}(M)=M$, and its restriction to the boundary then defines a boundary diffeomorphism $\phi=\left.\bar{\phi}\right|_M\,:\, M \rightarrow M$. At the level of vectorfields this translates to
\begin{equation}
\left.\bar g_\mathrm{o}(n,\bar \chi)\right|_M=0\quad\mbox{such that}\quad \chi=\left.\bar{\chi}\right|_M  \ .\label{bndchi}
\end{equation}
The condition on the left simply states that the normal part of the bulk vectorfield that generates the bulk diffeomorphism should vanish at the boundary, if the bulk diffeomorphism is to preserve the boundary.

In simplifying \eqref{wdw}, as a first step it is important to evaluate the momentum constraint \eqref{const1} on velocities of the form \eqref{veldef}. This leads to
\begin{equation}
\bar\nabla^i_{\!\!\mathrm{o}}\partial_{[i}\bar\chi_{j]}=0 \ . \label{momconst}
\end{equation}

Similarly, evaluating the Wheeler-de Witt metric $\bar{\mathrm{G}}$ on the velocities \eqref{veldef} one finds -- upon partial integration, which produces a bulk term that vanishes by the momentum constraint \eqref{momconst} so that only the boundary term remains -- that
\begin{equation}
\bar{\mathrm{G}}_{\phi^* \bar g_{\mathrm{o}}}(\phi^*(L_{\bar \chi_1}\bar g_{\mathrm{o}}),\phi^*(L_{\bar \chi_2}\bar g_{\mathrm{o}}))= \int_M \bar\chi_1^k n^j\bar\nabla_{\!\!\mathrm{o}(k}\bar\chi_{2\,j)} \sqrt{\det g}\,d^n x \ .
\end{equation}
One would like to interpret the above in terms of the boundary vector fields $\chi\in \mathfrak{X}(M)$ only. To do so one needs to determine the expression inside the integral in terms of $\chi$. First one observes, via a short computation\footnote{Along the boundary we split the tangent space as $\left.(\partial_i)\right|_M=\left.(n^i\partial_i,\partial_a)\right|_M$, so that a vector decomposes as $\left.(\bar v^i)\right|_M=(\left.\bar v^i\right|_M n_i,v^a)$. In this notation \eqref{bndchi} gets re-expressed as $\left.(\bar \chi^i)\right|_M=(0,\chi^a)$. See \cite{Kutluk:2019ghr} Appendix B for a detailed discussion.}, that 
\begin{equation}
 \left.\bar\chi^i_1 n^j \bar\nabla_{\!\!\mathrm{o}(i} \bar\chi_{2\,j)} \right|_{M} = \left.n^i\bar \chi_1^j \partial_{[i}\bar\chi_{2\,j]} \right|_{M} -2K_{ab}\chi_1^a\chi_2^b \ .\label{equality1}
\end{equation}
The second term, which includes the (pull-back of the) extrinsic curvature is already explicitly written in terms of boundary quantities, so it remains to determine $\partial_{[i}\bar\chi_{j]}$, in terms of $\chi$. Via \eqref{momconst} we can think of $\chi$  as providing a boundary condition for a boundary value differential problem for $\bar{\chi}$. If we define $\bar\theta=d\bar{\chi}$ then we see that \eqref{momconst} takes the form $\delta \bar{\theta}=0$ while simultaneously $d\bar\theta=0$. Here we recognize then the map introduced in (\ref{thetamap}, \ref{thetaeqs}) on our way to defining the operator $\mathbb{D}$ \eqref{Ddef}. Rephrased this way we thus get
\begin{equation}
 \left.\bar\chi_1^i n^j \bar\nabla_{\!\!\mathrm{o}(i} \bar\chi_{2\,j)} \right|_{M} = g_{ab}\chi_1^a (\mathbb{D}\chi_2)^b \ .
\end{equation}
In the end we thus find the following form for the Wheeler-de Witt metric $\bar{\mathrm{G}}$, pulled-back to the space of boundary diffeomorphisms: 
\begin{eqnarray}
\bar{\mathrm{G}}_{\phi^* \bar g_{\mathrm{o}}}(\phi^*(L_{\bar \chi_1}\bar g_{\mathrm{o}}),\phi^*(L_{\bar \chi_2}\bar g_{\mathrm{o}}))&=&\int_M g(\chi_1,\mathbb{D}\chi_2)\,\mathrm{vol}\,g=G^\mathbb{D}_{\phi}(\dot{\phi}_1,\dot{\phi}_2) \ . \label{GRmet}
\end{eqnarray}
Here we recognize the inner product \eqref{inprod} and the right invariant metric \eqref{Dmet} introduced in Section \ref{sec:diffg}. The dynamic Einstein equation \eqref{dynamicaleq} reduces to the geodesic equation obtained by varying \eqref{GRmet} in the adiabatic, or Manton, approximation.

Note that in our discussion so far we have ignored the Hamiltonian constraint \eqref{const2}. We'll discuss its role and interpretation in the next subsection.

\subsubsection{On the Hamiltonian constraint}
It is important to point out that in the derivation of the metric on the group of boundary diffeomorphisms from the Einstein equations the Hamiltonian constraint did not play any role. As we will now shortly discuss it has another, two-fold role to play however. 

First, one can easily see that the Hamiltonian constraint \eqref{const2}, evaluated on $\dot{ \bar g}$ satisfying the assumption \eqref{veldef} and restricted to $M$, implies that \eqref{GRmet} vanishes. In other words, it says that those $\chi$ related to an actual solution of the full Einstein equations will be {\it null} vectors on $\mathrm{Diff}(M)$, i.e. $(\chi,\mathbb{D}\chi)=0$. In particular then, the Hamiltonian constraint forces (adiabatic) solutions to Einstein's equations to be {\it null geodesics} on the group of boundary diffeomorphisms. That this is possible is due to the pseudo-Riemannian nature of the metric \eqref{GRmetric}. In \cite{Kutluk:2019ghr} it was shown that on the subspace of exact boundary vector fields\footnote{We say that a vector field is (co-)exact iff the dual one-form is (co-)exact.} the operator $\mathbb{D}$ is negative, while on another subspace of boundary vectorfields -- those for which the bulk extension $\bar \chi$ is co-exact -- $\mathbb{D}$ is positive. This implies that $\langle\cdot,\cdot\rangle=(\chi,\mathbb{D}\chi)$ has mixed signature and allows for null vectors\footnote{Note that zero eigenvectors of $\mathbb{D}$ would make $\langle\cdot,\cdot\rangle$ degenerate. It was argued in \cite{Kutluk:2019ghr} -- though not proven -- that the kernel of $\mathbb{D}$ corresponds to Killing vectors of the boundary metric. This would imply that $\langle\cdot,\cdot\rangle$ when restricted to the coset space $\mathrm{Diff}(M)/\mathrm{ISO}(M)$ is a non-degenerate, albeit pseudo-Riemannian, metric.}. 

Secondly, since the momentum constraint is only sensitive to $d\bar{\chi}^\flat$, it follows that it determines a bulk extension of the boundary vector field $\chi$ only up to an exact part. To make this more precise, one can split $\bar\chi^\flat$ in an exact and co-exact part:
\begin{equation}
\bar\chi^\flat=\bar\eta+d\bar\alpha\quad\mbox{where}\quad \delta\bar{\eta}=0 \ .\label{decomp}
\end{equation}
It was shown in \cite{Kutluk:2019ghr} that $\bar{\eta}$ is uniquely determined once $\chi$ , and hence $\bar\theta_{\chi^\flat}=d\bar{\chi}^\flat$, is given. One can then insert the decomposition \eqref{decomp} into the Hamiltonian constraint \eqref{const2} -- via \eqref{veldef} -- and it takes the form 
\begin{equation}
(\bar\nabla_i\partial^i\bar\alpha)^2-\bar\nabla_{i}\partial_{j}\bar\alpha \bar\nabla^i\partial^j\bar\alpha-2\bar\nabla^i\bar\eta^j\bar\nabla_{i}\partial_{j}\bar\alpha =\bar\nabla_{(i}\bar\eta_{j)} \bar\nabla^i\bar\eta^j\, .\label{alphaeq}
\end{equation}
Since $\bar\eta$ is already determined through the momentum constraint we can interpret \eqref{alphaeq} as a differential equation determining $\bar{\alpha}$, which should be solved under the boundary condition $d\left(\left.\alpha\right|_M\right)=\chi^\flat-\emb^*\bar\eta\,,\ \left. n^i\partial_i\alpha\right|_M=-\left.n^i\bar{\eta}_i\right|_{M}$, for a given boundary vector field $\chi$. Once $\bar \alpha$ is found and $\bar \chi^\flat$ is fully known one should integrate \eqref{veldef} to obtain the family of bulk diffeomorphisms $\bar{\phi}(t)$ and the metrics $\bar g(t)$, which then finally provide an $(n+2)$ dimensional metric \eqref{lormet} that (approximately) solves the vacuum Einstein equations.

Solving the Hamiltonian constraint in the form \eqref{alphaeq} remains however a daunting task, this equation is a non-homogeneous, non-linear PDE resembling the $(n+1)$-Hessian equation, see e.g. \cite{Wang:2009}. In the simple case $n=1$, see the next subsection, \eqref{alphaeq} can be recast in the form of the homogeneous real Monge-Ampere equation.

\subsection{The Case of 2 Spatial Dimensions}\label{2dsec}
In this subsection we focus on the special case $n=1$, i.e. $\dim M=1$ and $\dim \bar M=2$ and will provide a concrete form for the operator $\mathbb{D}$ defined in \eqref{Ddef}, the associated metric and geodesic equation. We will also show how the Hamiltonian constraint \eqref{alphaeq} can be reduced to the homogeneous real Monge-Ampere equation in this case. We first discuss a large class of examples in general, where the boundary is a polar curve, and afterwards make a few comments on the simplest cases, those of circle and ellipse.

\subsubsection{Generic polar curve}
\begin{figure}
	\begin{center}
	\includegraphics[scale=0.8]{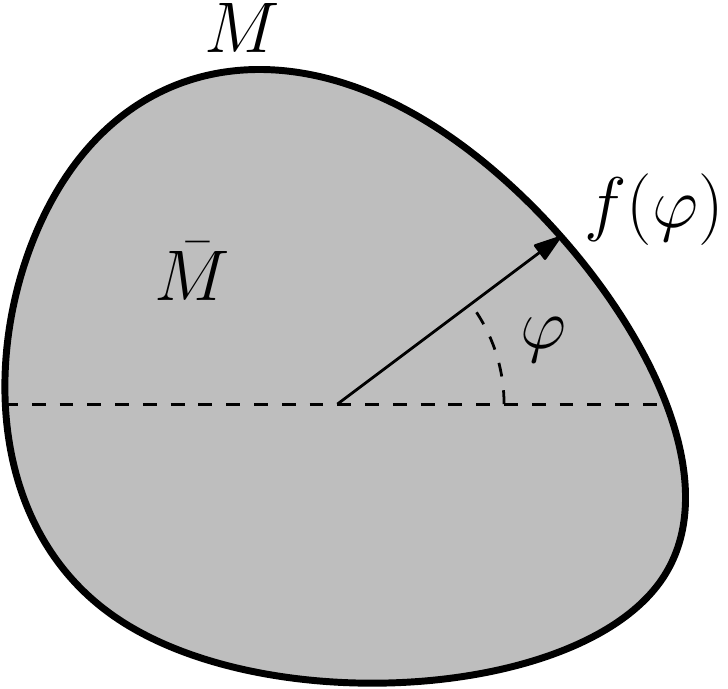}
\end{center}\caption{The polar curve $f(\varphi)$ determines the boundary $M$ of the region $\bar M$ in $\mathbb{R}^2$.}\label{2dfig}
\end{figure}

 Let us -- permitting a slight loss of generality --  concentrate on the following setup. We take $\bar M$ to be the simply connected region around the origin of the flat Eucidean plane, in the interior of the polar curve $r=f(\varphi)$, see Figure \ref{2dfig}, or explicitly
\begin{eqnarray}\label{M-Mbar}
\bar M&=&\{(r\cos\varphi, r\sin\varphi)\in \mathbb{R}^2\,|\, 0\leq r\leq f(\varphi)\,,\ \varphi\in [0,2\pi)\} \ , \\
M&=&\{(r\cos\varphi, r\sin\varphi)\in \mathbb{R}^2\,|\, r=f(\varphi)\,,\ \varphi\in [0,2\pi)\}  ,\\
\emb(\varphi) &=& \lp f(\varphi), \varphi \rp \ . \label{pullback}
\end{eqnarray}

Our first task is to compute the map $\bar{\theta}_\omega$, see \eqref{thetamap}. The boundary value problem \eqref{thetaeqs} needs to be slightly modified in the special case $n=1$, since the boundary condition $\emb^*\bar{\theta}_\omega=d\omega$ becomes contentless as both sides are trivially zero. To find an alternative boundary condition, let us recall the argument below \eqref{equality1}, which identifies the key relation 
\begin{equation}
\bar{\theta}_\omega=d\bar{\omega}\quad\mbox{and}\quad \emb^*(\bar\omega)=\omega \ .\label{thetadef1}
\end{equation}
Integrating \eqref{thetadef1} over $\bar M$, which is only possible when $n=1$ because then $\dim\bar M=\mathrm{deg}\, \bar{\theta}_\omega=2$, provides the condition $
 \int_{\bar M} \bar\theta_\omega= \int_M \omega
$.
Adding the information that we want $\bar{\theta}_\omega$ to solve the momentum constraint then leads to the replacement of definition \eqref{thetaeqs} in the special case $n=1$ by
\begin{equation}
\delta\bar\theta_\omega=0 \ , d\bar\theta_\omega=0 \gag \int_{\bar M} \bar\theta_\omega= \int_M \omega \sgd \label{2dbndc}
\end{equation}
Note that on $\bar M$ any two form that is co-closed is of the form $\bar{\theta}_\omega=\bar c\, \overline{\mathrm{vol}}$, with $\bar c$ a constant. Let us recall we work with the flat metric on $\bar M$, i.e.
\begin{equation}
d\bar{s}^2=dx^2+dy^2=dr^2+r^2d\varphi^2 \ ,\qquad ds^2=\cale_\varphi ^2d\varphi^2 \gwg \cale_\varphi=\sqrt{f'^2+f^2} \ ; 
\end{equation}
which define the unit volume forms
\begin{eqnarray}
\overline{\mathrm{vol}}&=& \,dx\wedge dy= \, r dr\wedge d\varphi \ , \\
\mathrm{vol}&=&\cale_\varphi d\varphi \ .\qquad 
\end{eqnarray}
Note that the area of $\bar M$ and its circumference -- i.e. the length of $M$ -- are given as
\begin{equation}
A=\int_{\bar M}\overline{\mathrm{vol}}=\frac{1}{2}\int_0^{2\pi} f^2 d\varphi \gag C=\int_{M}\mathrm{vol}=\int_0^{2\pi}\cale_\varphi d\varphi \ .
\end{equation}
It follows that we can solve \eqref{2dbndc} explicitly as
\begin{equation}
\bar{\theta}_\omega=p_\omega\, \overline{\mathrm{vol}} \gwg p_\omega=A^{-1}\int_M \omega \ . \label{pdef}
\end{equation}
Given the map $\bar{\theta}$, the definition \eqref{Ddef} of $\mathbb{D}$ remains unchanged. The period $p_\omega$ will appear through the {\it zeromode} of the vectorfield $\chi=\chi^\varphi \partial_\varphi$:
\begin{equation}
\pi(\chi)=p_{\chi^\flat}= A^{-1} \int_0^{2\pi} \cale_\varphi^2\,\chi^\varphi d\varphi \ . \label{pidef}
\end{equation}
To work out the action of $\mathbb{D}$ on $\chi$ explicitly, we need some information on the extrinsic geometry of $M$ as embedded in $\bar M$, in particular the normal vector field and extrinsic curvature
\begin{eqnarray}
n&=& f \cale_\varphi^{-1} \lp \partial_r - \frac{f'}{f^2} \partial_\varphi \rp  \ , \label{normvec}\\
K&=&\calk\, (\cale_\varphi d\varphi)^2\, \gwg \calk= \frac{f^2 + 2f'^2-f f''}{(f'^2+f^2)^{3/2}} \  .
\end{eqnarray}
Using these we then see the operator $\mathbb{D}$ acting on a vector field $\chi$ on $M$ is
\begin{align}
\mathbb{D} \chi &=  \pi(\chi) \cale_\varphi^{-1}\partial_\varphi - 2 \calk \chi \ ,\label{Delg}
\end{align}
or fully explicitly
\begin{equation}
\frac{1}{2}(\mathbb{D} \chi)^\varphi=\frac{\int_0^{2\pi}(f'^2+f'^2)\chi^\varphi d\tilde \varphi}{\sqrt{f'^2+f^2}\int_0^{2\pi}f^2d\tilde\varphi}-\frac{f^2 + 2f'^2-f f''}{(f'^2+f^2)^{3/2}}\chi^\varphi \ . \label{Dexpl}
\end{equation}
Once we have determined the operator $\mathbb{D}$, the metric \eqref{inprod} on $\mathfrak{diff}(M)=\mathfrak{X}(M)$ follows:
\begin{eqnarray}
\frac{1}{2}\langle \chi_1,\chi_2\rangle&=&\frac{1}{2}\int_0^{2\pi} \sqrt{g}\,g_{\varphi\varphi}\,\chi_1^\varphi(\mathbb{D}\chi_2)^\varphi d\varphi\\
&=&\left(\frac{A}{2}-\int_0^{2\pi}\cale_{\varphi}^{-1}\calf_{\varphi}^2\,\calk d\varphi\right)\pi(\chi_1)\pi(\chi_2)\nonumber\\
&&-\int_0^{2\pi}\cale_{\varphi}^{-1}\calf_{\varphi}\,\calk\,\big(\pi(\chi_1)\partial_\varphi \alpha_2+\pi(\chi_2)\partial_\varphi \alpha_1\big)d\varphi\label{metexp}\\
&&-\int_0^{2\pi}\cale_{\varphi}^{-1}\calk\,\partial_\varphi \alpha_1\partial_\varphi \alpha_2\,d\varphi \ . \nonumber
\end{eqnarray}
Here we used the following decomposition of the boundary vector field $\chi$:
\begin{equation}
\chi^\varphi= \cale_{\varphi}^{-2}\left(\pi(\chi) \calf_{\varphi}+\partial_\varphi \alpha\right)\ . \label{chidecomp}
\end{equation}
This decomposition is based on the Hodge decomposition of the dual one-form, $\chi^\flat=\frac{A}{C} \pi(\chi)\mathrm{vol}+d\alpha$ -- in which case $\calf_\varphi=\frac{A}{C} \cale_\varphi$ --, but can be more general. In particular, as we will see when discussing the Hamiltonian constraint, it can be useful to choose\footnote{Let us call $\bar\psi=\frac{1}{2}(xdy-ydx)=\frac{r^2}{2}d\varphi$, then $d\bar{\psi}=\overline{\mathrm{vol}}$ and thus $A=\int_{M} \emb^*\bar{\psi}$. But since that implies $\int_M(\frac{A}{C}\mathrm{vol}-\emb^*\bar{\psi})=0$ it follows that the difference between these two forms is exact. In other words, starting from the Hodge decomposition $\omega=\omega_0\,\frac{A}{C}\mathrm{vol}+d\alpha$ one can obtain the alternative decomposition $\omega=\omega_0\, \emb^*\bar{\psi}+d\tilde \alpha$.} $ \chi^\flat=\frac{1}{2}\pi(\chi)\emb^*(xdy-ydx)+d\alpha$, in which case
\begin{equation}\label{eq:F-choice-2}
\calf_\varphi=\frac{f^2}{2} \ .
\end{equation} 
We will always make this choice for the decomposition of the boundary vectorfield $\chi$ in the following. Under the decomposition \eqref{chidecomp}  we have split the vector field $\chi$ into its zeromode $\pi(\chi)$, which is a real number, and a (periodic) function $\alpha(\varphi)$.

The explicit form of the metric \eqref{metexp} shows that indeed the operator $\mathbb{D}$ is symmetric, and that in general the metric can have mixed signature.

The geodesic equation on $\mathrm{Diff}(M)$ associated to this metric can be worked out via the generic form \eqref{geoeq_expl}:
\begin{equation}
\partial_t \chi^\varphi+3\chi^\varphi \nabla_\varphi \chi^\varphi+(\chi^\varphi)^2\partial_\varphi\log \calk=(\cale_\varphi \calk)^{-1}\left(\frac{1}{2}\pi(\partial_t\chi)+\nabla_\varphi \chi^\varphi\,\pi(\chi)\right)\label{2dgeo}
\end{equation}
where
\begin{eqnarray}
\nabla_\varphi \chi^\varphi&=&\partial_\varphi \chi^\varphi+\Gamma_{\varphi\varphi}^\varphi \chi^\varphi =\partial_\varphi \chi^\varphi +\chi^\varphi \partial_\varphi\log \cale_\varphi \ .
\end{eqnarray}

In the first two terms on the left hand side of \eqref{2dgeo} we recognize Burger's equation \eqref{burgeq}, but we see there are additional terms, in particular the right hand side where an integral operator on $\chi$ appears.

Finally let us comment on the Hamiltonian constraint in this class of examples. In the decomposition \eqref{decomp} we take
\begin{equation}
\bar \eta =\pi(\chi)  \frac{r^2}{2} d\varphi\label{eta}\,,
\end{equation}
so that indeed $d\bar{\eta}=\bar\theta_{\chi^\flat}$. 
Note that $\bar\nabla_{(i}\bar{\eta}_{j)}=0$, reflecting the fact that the vectorfield $\bar \eta^i$ is Killing, indeed it generates rotations around the origin in $\bar M$. It follows that the contribution of $\bar\eta$ to the Hamiltonian constraint \eqref{alphaeq} vanishes so that it simplifies to
\begin{equation}
(\bar\nabla_i\partial^i\bar\alpha)^2-\bar\nabla_{i}\partial_{j}\bar\alpha \bar\nabla^i\partial^j\bar\alpha=0 \ .
\end{equation}
In Cartesian coordinates on $\bar M$, this reduces to the homogeneous real Monge-Ampere (HRMA) equation:
\begin{equation}
\label{eq:MA}
\lp \pr_x \pr_y \bar \alpha \rp^2- \pr_x^2 \bar \alpha \pr_y^2 \bar \alpha =0 \sgd
\end{equation}
The Monge-Ampere equation is a well-studied non-linear PDE that has its historical origins in the geometric problem to find a hypersurface with a prescribed extrinsic curvature. I.e. one interpretation of \eqref{eq:MA} is that a two dimensional surface with the embedding $\lp x,y, \bar\alpha(x,y) \rp$ in $\R^3$ has zero extrinsic curvature. This observation is probably not all too relevant to our setup, but we will use in an example below one of the (few) explicit solutions\footnote{See \cite{polyanin} for an overview.} to this equation that are known. 

In addition to solving \eqref{eq:MA}, note that $\bar\alpha$ needs to satisfy certain boundary conditions: these can be extracted from the conditions $\emb^* \bar{\chi}=\chi$ and $\left.n^i\bar{\chi}_i\right|_{M}=0$, which applied to \eqref{decomp} -- together with (\ref{pullback}, \ref{normvec}, \ref{chidecomp}, \ref{eq:F-choice-2}, \ref{eta}) provide
\begin{equation}
\left.\bar{\alpha}\right|_M=\alpha \gag \left. n^i \partial_i \bar{\alpha}\right|_M=\frac{f f'}{2\sqrt{f'^2+f^2}}\pi(\chi) \ \label{boundcond}
\end{equation}

Interestingly the boundary conditions \eqref{boundcond} are of Cauchy type, rather than of the typical boundary value type used for elliptic equations. Still this is not necessarily a problem, due to the pseudo-elliptic nature of the HRMA equation, although it is not obvious that the solution will be free of singularities.

\subsubsection{Circle}
The simplest special case is clearly that of the circle $M=S^1$, where $f(\varphi)=R$ and $\calk=R^{-1}$. One finds that
\begin{equation}
\mathbb{D}\chi=R^{-1}\left(\pi(\chi)\partial_\varphi-2\chi\right)=-2R^{-3}\partial_\varphi \alpha\, \partial_\varphi \ .
\end{equation}

The rightmost expression -- which is obtained upon using the decomposition \eqref{chidecomp} -- shows that the zeromode , $\pi(\chi)=\frac{1}{\pi}\int_0^{2\pi}\chi^\varphi d\varphi$, spans the kernel of $\mathbb{D}$. This is no surprise, since such a zeromode describes a vector field of the form $\chi= c \partial_\varphi$, which in this special case is a Killing vector of the boundary. This is in accord with the conjecture of \cite{Kutluk:2019ghr} that the kernel of $\mathbb{D}$ corresponds to isometries of the boundary. 

The inner product \eqref{metexp} becomes
\begin{equation}
\langle \chi_1,\chi_2\rangle=-2 R^{-2}\int_0^{2\pi}\partial_\varphi\alpha_1\partial_\varphi\alpha_2\, d\varphi \ .\label{S1met}
\end{equation}
Since the zeromode $\pi(\chi)$ has disappeared, the corresponding metric on $\mathrm{Diff}(S^1)$ is degenerate. Instead it has a natural interpretation as a right invariant, non-degenerate metric on  $\mathrm{Diff}(S^1)/\mathrm{SO}(2)$, since the zeromode, which can now be consistently quotiented out, generates the rotations of the circle.

From the point of view of GR this example is not very relevant, since the metric \eqref{S1met} is negative definite and will thus not allow for null geodesics. I.e. there will be no solutions to the Hamiltonian constraint \eqref{eq:MA}, since the compatibility of the boundary conditions \eqref{2dbndc} with the differential equation is exactly the condition of vanishing norm. From a broader mathematical perspective one can of course consider the geodesics irrespective of a condition on their tangent vectors. The geodesic equation \eqref{2dgeo} for the metric \eqref{S1met} reduces to Burger's equation for the function $u=\partial_\varphi \alpha$.

\subsubsection{Ellipse}
To provide an example where there are null directions, so that it is of relevance to GR, one can consider the case where the boundary is an ellipse, described by the polar curve
\begin{equation}
f(\varphi)= \frac{\rho}{2} \frac{\sinh(2 \U)}{\sqrt{\cosh^2\U-\cos^2\varphi}} \ .
\end{equation}
Here the positive constants $\rho$ and $\U$ set the scale and eccentricity $e=1/\cosh\U$, respectively. It is computationally convenient to replace the polar coordinates we have been using by elliptic coordinates for this example: 
\begin{align}
x&=\rho \cosh u \cos v \ , \\
y&=\rho \sinh u \sin v \ ,
\end{align}
where $u \geq 0$ and $2 \pi > v \geq 0$. In these coordinates\,{}\footnote{Note that the angular coordinate $v$ along the boundary is related to the one we have used previously as
\begin{equation*}
v=\arctan\left(\frac{\tan\varphi}{ \tanh \U}\right) \ .
\end{equation*}
}	 the boundary ellipse is simply the curve $u=\U$, while the flat metric on $\bar M$ and the induced metric on $M$ take the form
\begin{equation}
d\bar s^2=\rho^2 (\cosh^2u-\cos^2v) (du^2+dv^2) \gag ds^2=\cale_v^2 dv^2 \gwg \cale_v=\rho\sqrt{\cosh^2\U-\cos^2 v} \ .
\end{equation}
Furthermore one has 
\begin{eqnarray}
n=\cale_v^{-1}\partial_u \ ,\qquad K=\calk (\cale_vdv)^2 \ ,\qquad \calk=\frac{\rho^2}{2}\cale_v^{-3}\sinh 2\U \ , \qquad \calf_v=\frac{\rho^2}{4}\sinh 2\Upsilon \ .
\end{eqnarray}
A short computation then provides
\begin{eqnarray}
\frac{1}{2}\langle\chi_1,\chi_2 \rangle&=&-\frac{\pi\rho^2}{4}e^{-2\Upsilon}\pi(\chi_1)\pi(\chi_1)\nonumber\\
&&-\frac{\sinh^2 2\Upsilon}{8}\int_0^{2\pi}\frac{\pi(\chi_1)\partial_v\alpha_2+\pi(\chi_2)\partial_v\alpha_1}{(\cosh^2 \Upsilon-\cos^2 v)^2} d v\label{metellips}\\
&&-\frac{\sinh 2\Upsilon}{2\rho^2}\int_0^{2\pi}\frac{\partial_v\alpha_1\partial_v\alpha_2}{(\cosh^2\Upsilon-\cos^2 v)^2} d v\nonumber
\end{eqnarray}
where
\begin{equation}
\pi(\chi)=\frac{2}{\pi \sinh 2\Upsilon}\int_0^{2\pi} (\cosh^2\U-\cos^2 v)\chi^v dv \ .
\end{equation}
The geodesic equation \eqref{geoeq_expl} takes the form
\begin{equation}
\sinh 2\Upsilon(\partial_t \chi^v+3\chi^v \partial_v \chi^v)-\sin 2v\, \chi^v \pi(\chi)=(\cosh^2\Upsilon-\cos^2v)\left(\pi(\partial_t\chi)+2\partial_v\chi^v\pi(\chi)\right) \ . 
\end{equation}

Although one could further work out the metric -- and possibly the geodesic equation as well -- by applying a Fourier decomposition to $\alpha$, we will refrain from providing details since the answer appears -- to us -- not particularly enlightening.

As we have pointed out before, any solution to the Hamiltonian constraint -- equivalent here to the HRMA \eqref{eq:MA} -- restricted to the boundary, provides a null direction for the metric \eqref{metellips}. Interestingly on the ellipse we managed to find a closed form solution of the HRMA satisfying the correct boundary conditions. Given that
\begin{equation}
\bar \eta=\pi(\chi)\frac{\rho^2}{4}(\sin 2v\, du+\sinh 2u\, dv)
\end{equation}
one finds that the boundary conditions for the HRMA in terms of $\chi$ become 
\begin{equation}
\left. \pr_u \bar{\alpha} \rli_{u=\U}= - \frac{\rho^2\,\pi(\chi)}{4} \sin(2v) \gag \left. \bar{\alpha} \rli_{u=\U}=  \alpha \ ,\label{ellipsbndcond}
\end{equation}
where we have decomposed the boundary vectorfield as in \eqref{chidecomp}. The natural problem is to find a solution $\bar{\alpha}$ to HRMA satisfying \eqref{ellipsbndcond} given a boundary vector field $\chi$ that is null with respect to \eqref{metellips}. But one can also reverse the logic: any solution $\bar \alpha$ to HRMA satisfying the left condition of \eqref{ellipsbndcond} will provide a boundary vectorfield via the right equation of \eqref{ellipsbndcond} that is null with respect to \eqref{metellips} -- see \cite{Kutluk:2019ghr} for an argument.

\begin{figure}
	\begin{center}
	\includegraphics[scale=0.5]{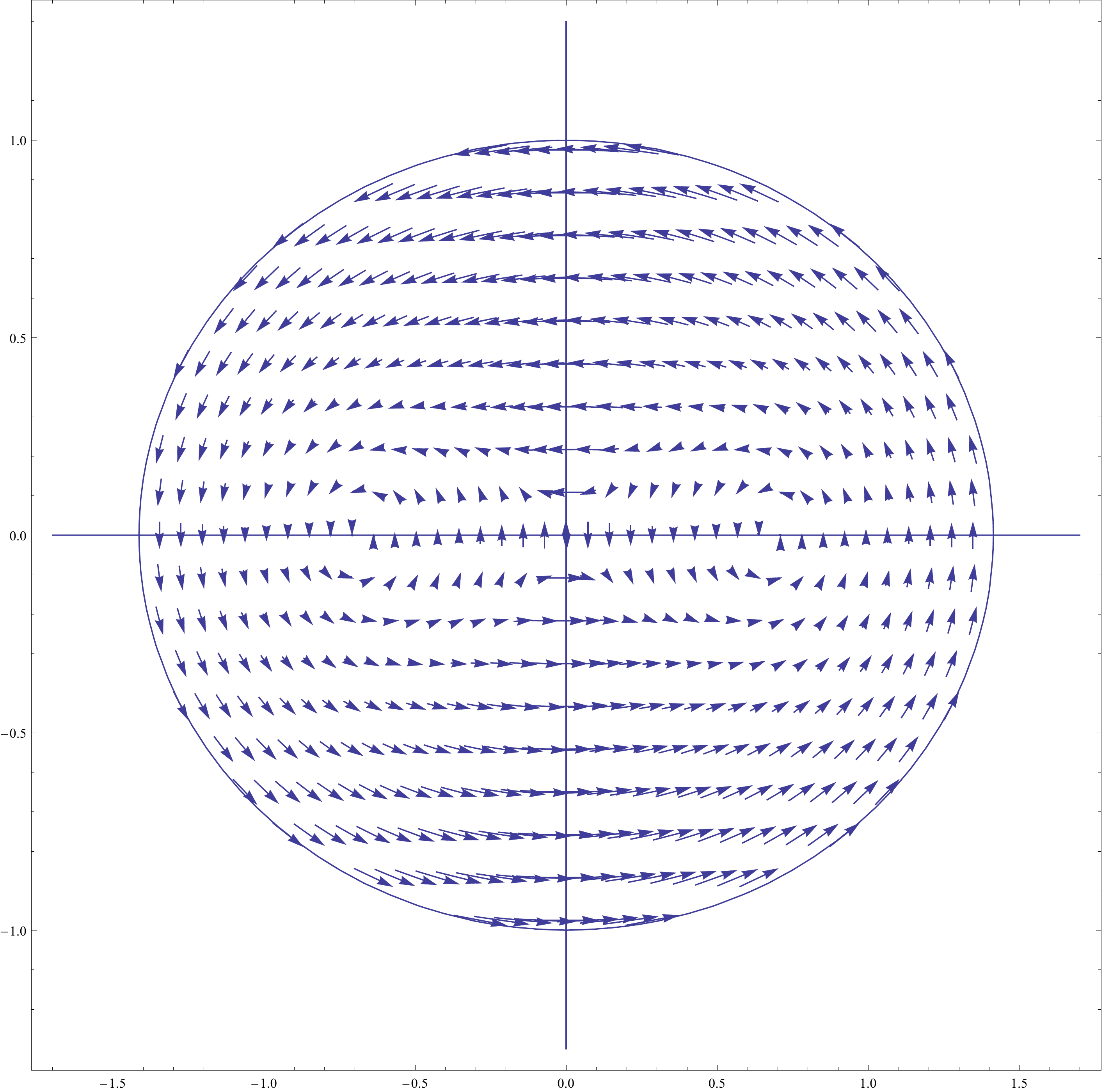}
	\caption{The bulk vector field $\bar \chi$ associated to \eqref{solexp}, when $\cosh\U=\sqrt{2}$ and $\rho^2\pi(\chi)=1$.}
	\label{fig}
	\end{center}
\end{figure}

One can verify that the following function solves \eqref{eq:MA}\footnote{This solution is within the class of solutions to HMRA that are of the form \cite{polyanin}
\begin{equation*}
\bar\alpha(x,y)= y \ \Phi \lp \frac{y}{x} \rp \ .
\end{equation*}
} and satisfies the left equality of \eqref{ellipsbndcond}:
\begin{equation}
\bar \alpha (x,y)= -\frac{\rho^2\,\pi(\chi)}{2 \cosh \U} \sgn(x) \, y \,\, _2F_1\lp \frac{1}{2}, \frac{\cosh^2 \U}{2};\frac{\cosh^2 \U}{2}+1;- \frac{1}{\tanh^2 \U} \frac{y^2}{x^2} \rp \ .\label{solexp}
\end{equation}
The corresponding vectorfield $\bar{\chi}=(\bar{\eta}+d\bar{\alpha})^\sharp$ is plotted in Figure \ref{fig} in the special case\footnote{If $\cosh \U= \sqrt{2} $ then
	\begin{equation*}
	_2F_1\lp \frac{1}{2}, 1,2,- 2\frac{y^2}{x^2} \rp =  \frac{x^2}{y^2} \lp -1 + \sqrt{1+ 2 \frac{y^2}{x^2}} \rp \sgd
	\end{equation*}} where $\cosh\U=\sqrt{2}$ and for $\rho^2\pi(\chi)=1$. 

It is important to point out that this vector field is discontinuous at the origin, which means that the corresponding solution to GR will have a singularity there. We'll comment further on this observation and possible consequences in the discussion section.

\section{Discussion}\label{sec:discuss}

We reviewed how the $n+2$ dimensional Einstein equations give rise to geodesic motion on the group of diffeomorphisms of the $n$ dimensional boundary $M$ of a finite $n+1$ dimensional spatial region $\bar M$, as in \cite{Kutluk:2019ghr}. The metric one obtains on $\mathrm{Diff}(M)$ can be interpreted as the pull-back of the Wheeler-deWitt metric onto an  embedding of $\mathrm{Diff}(M)$ in superspace. It takes the standard form $G_{\phi}(\dot{\phi}_1,\dot{\phi}_2)=\int_M g(dR_{\phi^{-1}}\dot{\phi_1},\mathbb{D}\, dR_{\phi^{-1}}\dot{\phi_2})\mathrm{vol}\, g$ of a right-invariant metric, with $\mathbb{D}$ a symmetric operator on the space of vector fields $\mathfrak{X}(M)=\mathfrak{diff}(M)$. The geodesic equation for some simple choices for $\mathbb{D}$ correspond to some well known equations of continuum mechanics, as we reviewed in Section \ref{sec:ex}. In \eqref{Ddef} we provided a new, more elegant, form for the operator $\mathbb{D}$ obtained from the Einstein equations. This operator differs from the standard examples in continuum mechanics in a number of ways. First of all it is not determined in terms of data intrinsic to $M$, but rather uses properties that are extrinsic to $M$ and determined in terms of its embedding as the boundary of $\bar M$. Secondly the operator $\mathbb{D}$ obtained in this way is not positive, so that the associated metric $G$ on $\mathrm{Diff}(M)$ is pseudo-Riemannian rather than Riemannian. Finally we should point out that in this case $\mathbb{D}$ is an integral operator rather than a differential operator. This has the consequence that the associated geodesic equation is an integro-differential equation rather than just a PDE such as in the standard continuum mechanics examples. These features were illustrated in detail in the $n=1$ class of examples discussed in Section \ref{2dsec} .

The geodesic motion of our setup describes a one parameter family of boundary diffeomorphisms; to reconstruct a $n+2$ dimensional Lorentzian metric out of this, one first of all needs to extend this boundary diffeomorphism into the bulk space $\bar M$. This goes in two steps. The co-exact part of the vector field generating the bulk diffeomorphism is uniquely determined through the momentum constraint of GR; this is a linear equation which is well under control. The exact part however can only be obtained upon solution of the Hamiltonian constraint, which is a complicated non-linear PDE of Hessian type. As we showed in Section \ref{2dsec}, in the case $n=1$ the Hamiltonian constraint becomes the homeogeneous real Monge-Ampere (HRMA) equation. This equation should be solved under the Cauchy type boundary conditions \eqref{boundcond}, which puts into doubt the existence of a smooth solution. Indeed, in the special case of an elliptical boundary we managed to find the explicit closed form solution \eqref{solexp} which leads to a mildly singular -- discontinuous at a point -- bulk vector field. As a consequence the corresponding time dependent Lorentzian metric can be expected to have a point singularity as well. Obtaining such a metric explicitly remains out of reach however, as -- even in this simple case -- we could only solve the HRMA equation for a very special boundary condition, not for a one-parameter family of boundary conditions corresponding to an actual geodesic on $\mathrm{Diff}(M)$.  In general the three step procedure of
\begin{itemize}
	\item [1)] solving the geodesic equation on $\mathrm{Diff}(M)$ for the boundary vectorfield
	\item[2)] solving the Hamiltonian constraint to obtain the bulk extension of the vectorfield
	\item [3)] integrating the bulk vector field to a bulk diffeomorphism
\end{itemize}
appears too challenging to carry out explicitly. It could be interesting to explore this numerically. 

So far we have only considered the setup of a compact spatial region with boundary. An interesting alternative would be to consider a non-compact spatial region with an interior boundary, or possibly even a non compact region with only an asymptotic boundary. Furthermore one could consider the inclusion of a cosmological constant or other sources.  Although we expect our method to remain largely unchanged and applicable in such generalizations, this has not been explored.

Finally let us shortly comment on our work in the context of the general theory of gauge symmetry in the presence of boundaries, a topic which has seen much recent activity -- see e.g. \cite{Donnelly:2014fua, Donnelly:2016auv, Speranza:2017gxd, Blommaert:2018oue, Harlow:2019yfa, Barnich:2019qex, Riello:2019tad, Geiller:2019bti, Riello:2020zbk, Donnelly:2020xgu}. The presence of a boundary affects the {\it phase space}, see \cite{Riello:2021lfl} for a recent in depth discussion and further references. Such work tries to understand in detail the kinematics -- i.e. symplectic structure -- of gauge theories in this setting.  The work \cite{Kutluk:2019ghr} on which this paper is based, as well as the earlier results \cite{Lechtenfeld:2015uka, Seraj:2017rzw} in Yang-Mills theory -- see also \cite{Cork:2021paa} -- are rather a first attempt to gain an understanding of the dynamics -- i.e. Hamiltonian -- of that sector. The key point is that the extra degrees of freedom associated to the boundary also cover a new region of {\it configuration space} equipped with a non-trivial metric \cite{Wheeler:1957mu, Babelon:1980uj} describing the dynamics. Actually the global gauge sector\footnote{With `global gauge sector' we mean those gauge orbits in configuration space which are not quotiented out, these are typically those generated by gauge transformations that are non-trivial on the boundary.} of configuration space is in some sense the simplest one, -- at least when restricted to orbits of simple configurations such as the vacuum -- as it has by definition a homogeneous space structure. It is also in this sense that the work discussed here is quite similar to the continuum mechanics setting, where the diffeomorphism group plays the role of the configuration space of the continuum material.

\section*{Acknowledgements}
We thank I. Gahramanov, B. G\"urel, B. Oblak and A. Seraj for useful discussions. EŞK is supported by the Research Fund of the Middle East Technical University, Project Number DOSAP-B-105-2021-10763. DVdB is partially supported by the Boğaziçi University Research Fund under grant number 21BP2.
\appendix
\section{Translation of notation}\label{notapp}

\begin{eqnarray*}
\mbox{Notation of \cite{Kutluk:2019ghr}}&\qquad& \mbox{Notation here}\\
M &\rightarrow & \bar M\\
\partial M &\rightarrow & M\\
h_{\mathrm{o}\,} & \rightarrow & \bar g_{\mathrm{o}}\\
h & \rightarrow & \bar g\\
k&\rightarrow & g\\
\chi & \rightarrow & \bar{\chi}\\
\zeta & \rightarrow & \chi\\
D^\perp \chi & \rightarrow & \lp j^*(\imath_n \bar \theta_{\chi^\flat})-\imath_{\chi}K \rp^\sharp \\
\eta&\rightarrow& \bar \eta\\
\phi & \rightarrow& \bar \alpha
\end{eqnarray*}

\bibliographystyle{utphys}
\bibliography{geo_lit}
\end{document}